\documentclass[%
reprint,
amsmath,amssymb,
aps,
superscriptaddress, 
nolongbibliography
]{revtex4-2}

\usepackage{graphicx}
\usepackage{dcolumn}
\usepackage{bm}
\usepackage{placeins}
\usepackage{float}
\usepackage{caption}
\usepackage{chemformula}

\begin{document}
	
\preprint{APS/123-QED}

\title{The influence of film thickness and uniaxial anisotropy on in-plane skyrmions: Numerical investigations of the phase space of chiral magnets}

\author{C. Rudderham}
\affiliation{Department of Physics and Atmospheric Science, Dalhousie University, Halifax, Nova Scotia, Canada B3H 3J5}

\author{M. L. Plumer}
\affiliation{Department of Physics and Atmospheric Science, Dalhousie University, Halifax, Nova Scotia, Canada B3H 3J5}
\affiliation{Department of Physics and Physical Oceanography, Memorial University of Newfoundland, St. John's, Newfoundland, Canada A1B 3X7}

\author{T. L. Monchesky}
\affiliation{Department of Physics and Atmospheric Science, Dalhousie University, Halifax, Nova Scotia, Canada B3H 3J5}%

\date{\today}

\begin{abstract}
The equilibrium phase space of magnetic textures in thin-films of cubic chiral ferromagnets, including skyrmions, is explored as a function of in-plane magnetic field strength, film thickness and uniaxial anisotropy. The interplay between these system parameters is found to give rise to a phase space with a rich structure, distinct from that of the nano-stripes that have been previously studied. For certain values of the anisotropy, the range of thicknesses supporting in-plane skyrmions and/or helicoids with an out-of-plane propagation vector is found to be disconnected, suggesting a possible direction for future experiments. We explain how this interesting phase space topology arises due to the geometric confinement of the thin-film system, and identify the optimal parameter ranges for future explorations of novel magnetic textures such as the oblique spiral phase. 

\end{abstract}

\maketitle

\section{\label{sec:level1}Introduction}

Cubic chiral ferromagnets are a class of magnetic materials that possess crystal structures that are cubic, but non-centrosymmetric. These include materials such as the intermetallic B20 compounds MnSi, MnGe and FeGe, as well as $\beta$-Mn CoZnMn alloys, and the Mott insulator \ch{Cu2OSeO3}. The lack of inversion symmetry in these materials, combined with spin-orbit interactions, gives rise to an anti-symmetric exchange interaction, also known as the Dzyaloshinskii-Moriya interaction (DMI)\cite{Dzyaloshinsky1958, Moriya1960a, Moriya1960b}. The interplay between the DMI and more conventional magnetic interactions such as the ferromagnetic exchange and Zeeman interactions results in the stabilization of a rich variety of magnetic phases, including quantized helicoidal states\cite{Wilson2013}, oblique spiral states\cite{Leonov2020b}, surface spirals\cite{Rybakov2016, Turnbull2022}, chiral bobbers\cite{Rybakov2015, Ahmed2018, Ran2021, Zheng2018} and perhaps most importantly, magnetic skyrmions, which are localized chiral magnetic textures that are of interest not just for their novel topological properties, but also for their possible applications in future magnetic storage technologies\cite{Zhang2015, Zhang2020a}. \newline

\noindent
Although first proposed theoretically in 1989\cite{Bogdanov1989}, the first experimental evidence for skyrmions was not obtained until 2009, when Muhlbauer et. al. demonstrated that there exists a small range of temperatures and applied fields for which bulk MnSi supports a skyrmion phase\cite{Muhlbauer2009}. This discovery kicked off a flurry of research into the magnetic properties of MnSi\cite{Pappas2009, Adams2011}, as well as other chiral cubic ferromagnets such as FeGe\cite{Wilhelm2011, Wilhelm2012}, MnGe\cite{Tanigaki2015, Kanazawa2012}, $\beta$-Mn CoZnMn alloys\cite{Henderson2023, Henderson2021, Karube2020, Karube2016}, and \ch{Cu2OSeO3}\cite{Seki2012a, Seki2012b, Adams2012, Qian2018, Chacon2018}. In particular, there has been extensive research into the how the magnetic properties of these materials change when prepared as thin samples, e.g. as epitaxially grown films\cite{Ahmed2018, Karhu2011, Wilson2012, Huang2012, Li2013, Meynell2014a, Meynell2014b, Sinha2014, Wilson2014, Porter2015, Yokouchi2015, Meynell2017} or mechanically polished lamellae\cite{Turnbull2022, Tonomura2012, Birch2020, Wilson2020}. \newline 

\noindent 
The geometric confinement introduced in thin chiral ferromagnetic systems can have significant effects on the properties of their magnetic phases, and in particular on the properties of the skyrmion phases. These confinement effects, however, are qualitatively different depending on the orientation of the external magnetic field with respect to the sample. If the external magnetic field is applied along the sample’s shortest axis, i.e., along the out-of-plane direction, then one obtains an out-of-plane skyrmion lattice phase. In this configuration, the skyrmion structures acquire additional chiral modulations near the film surfaces that serve to lower their energy\cite{Meynell2014b, Leonov2016a}, stabilizing the skyrmion phase over a larger range of temperatures and fields than in the bulk. This increased stability, combined with the relative ease with which the phase can be imaged directly using techniques such as Lorentz transmission electron microscopy (LTEM)\cite{Yu2011, Yu2010b, Zhao2016, Zheng2021, Meynell2023}, makes the out-of-plane configuration arguably the most well-studied of the skyrmion phases. 
\newline 

\noindent
If instead the external field is applied \textit{perpendicular} to the sample’s shortest axis, i.e., along the in-plane direction, then one can obtain an in-plane skyrmion lattice phase\cite{Yokouchi2015, Meynell2017}. Unlike their out-of-plane counterparts, these in-plane skyrmion structures do not make contact with the upper and lower film surfaces, and hence do not have their energy lowered by the aforementioned surface modulations. The film thickness, however, can still have a significant effect on the energetics of the in-plane skyrmion phase via its effect on properties such as the skyrmion packing density. Interestingly, this geometric confinement can impart a discretization to different magnetic phases, notably helicoid states with and out-of-plane helicoidal vector and a discrete number of twists\cite{Wilson2013, Kanazawa2016}, or in-plane skyrmion lattices with a discrete number of layers\cite{Yu2020}. In principle, therefore, the film thickness can be used as an easily adjustable control parameter for adjusting the stability of various magnetic phases. \newline

\noindent
One productive avenue of research has been to study the properties of magnetic nanostripes\cite{Jin2017, Du2015a, Song2018} or nanowires.\cite{Du2014, Du2015b, Liang2015, Yu2013}  The results obtained in these investigations, however, are not directly applicable to wider thin films, due to the importance of edge effects in nanostripe/nanowire geometries. The geometric confinement along the applied field direction produces energy-lowering edge modulations in the skyrmion structures analogous to what occurs near the sample surfaces for out-of-plane skyrmions. \newline

\noindent 
True in-plane skyrmion lattice phases, i.e., in-plane lattices obtained in thin-films wide enough that edge effects can be ignored, have not received as much attention as their out-of-plane or nanostripe counterparts. In thin-film MnSi, for instance, only two studies have provided experimental evidence of an in-plane skyrmion lattice phase\cite{Yokouchi2015, Meynell2017}. Film thickness is seemingly an important factor, with studies on thicker films of MnSi showing no evidence of an in-plane skyrmion lattice phase.\cite{Wiedemann2017}. Curiously, an in-plane skyrmion lattice phase has yet to observed in thin films of FeGe, a system whose mean-field magnetic properties are believed to be qualitatively similar to those of MnSi, although skyrmions have been observed near the sample edges\cite{Birch2020}. \newline

\noindent
Studying the properties of thin-film chiral ferromagnets also requires consideration of the effects of magnetocrystaline anisotropy. While cubic chiral ferromagnets are known to possess an intrinsic  cubic anisotropy\cite{ Chacon2018, Preissinger2021, Adams2018}, in epitaxial thin films its effects are often dominated by a much stronger uniaxial anisotropy. When grown on top of a semiconducting substrate, the lattice mismatch between the chiral ferromagnet and the substrate induces a strain in the magnetic material, which in turn induces a uniaxial magnetocrystalline anisotropy along the out-of-plane direction\cite{Karhu2012}. Depending on whether the induced strain is compressive or tensile, the resulting anisotropy can be either easy-axis (which helps stabilize out-of-plane skyrmions\cite{Huang2012, Butenko2010}), or easy-plane (which stabilizes in-plane skyrmions)\cite{Wilson2012}. In principle, the value of this uniaxial anisotropy can be tuned via the choice of substrate. Semiconducting substrates in particular are of interest because of the possibility of controlling skyrmions with electric fields\cite{Upadhyaya2015, Hsu2017} or currents\cite{Tchoe2012, Iwasaki2013}, and/or interfacing with existing technologies. \newline

\noindent
As a consequence of these considerations, the stability and properties of the various magnetic phases supported by thin-film chiral ferromagnets are expected to depend sensitively on the values of the applied in-plane field $H$, the uniaxial anisotropy $K_u$, and the film thickness $t$. This results in a rich three-dimensional phase space of parameters that can be controlled by experimentalists, at least in principle.  This phase space, however, remains relatively unexplored, even theoretically. There have been numerical investigations of the $(H, K_u)$ dependence of phases in an infinite thin-film of fixed thickness \cite{Leonov2020b}, and of the $(H, t)$ dependence of phases in an isotropic $(K_u=0)$ nanostripe\cite{Jin2017}, but we are unaware of any studies exploring the $(H, t)$ dependence of the phases supported by infinite thin-films, with or without uniaxial anisotropy. Nevertheless, such investigations could prove extremely useful, both by guiding experimentalists towards the most promising regions of phase space for future investigations and/or by explaining existing experimental data. \newline

\noindent
To this end, in this work we employ numerical micromagnetics to generate a series of equilbrium phase diagrams in $(H,t)$ space for infinite thin-film chiral ferromagnets with varying amounts of uniaxial anisotropy, $K_u$. We find that the equlibrium phase space of infinite thin film geometries is distinctly different from that of finite nanostrip geometries, and show that for certain uniaxial anisotropy values the range of film thicknesses that support stable skyrmion and/or helicoid phases can be discontinuous. Our results provide insight into the interplay between the various magnetic interactions present in these systems, and will help experimentalists determine the most promising system parameters to target in future film growths.

\section{Model and Methods}

\noindent
We consider thin-film systems of chiral ferromagnets that are infinite along the $x$ and $y$ directions, but possess a finite thickness $t$ along the $\hat{z}$ direction. This limit allows us to investigate the properties of films that are wide enough that edge effects have a negligible effect on the system's properties, while still maintaining the effects of geometric confinement along the out-of-plane direction. We assume that the magnetization $\mathbf{M}$ has a uniform magnitude $M_{sat}$ everywhere within the film, allowing us to work with the unit vector field $\mathbf{m}$, defined such that $\mathbf{M} = M_{sat} \mathbf{m}$. In the absence of thermal fluctuations, the standard micromagnetic energy density function for such systems contains four distinct terms\cite{Karhu2012, Wilson2012}, each corresponding to a different magnetic interaction, as shown in Equation \ref{hamiltonian}:

\begin{equation}
	w = A(\nabla \mathbf{m})^2 + D \mathbf{m} \cdot (\nabla \times \mathbf{m}) - \mu_0 M_{sat} \mathbf{m}\cdot \mathbf{H} - K_u m_z^2. 
	\label{hamiltonian}
\end{equation}

\noindent
The four relevant interactions are ferromagnetic exchange (with constant $A$), the Dzyaloshinskii-Moriya interaction (with constant $D$), the Zeeman interaction (with external magnetic field $\mathbf{H}$), and a uniaxial magnetocrystaline anisotropy along the $z$-axis (with constant $K_u$). In what follows, we consider systems where the external magnetic field is applied in-plane, specifically along the $\hat{x}$ direction. A negative value of $K_u$ corresponds to a hard-axis (easy-plane) anisotropy, while a positive $K_u$ corresponds to an easy-axis (hard-plane) anisotropy.
The material parameters present in Equation \ref{hamiltonian} can be used to define a characteristic length scale $L_D = 4\pi \frac{A}{D}$, as well as characteristic values for the field strength, $\mu_0 H_D = \frac{D^2}{2AM_{sat}}$, anisotropy,  $k_u = K_u A /D^2$, and volumetric energy density, $\omega_0 = D^2 /2A$. Our results are presented in terms of these quantities, so that they may be applied to all systems described by Equation \ref{hamiltonian}, regardless of the specific values of the material-dependent parameters.  \newline

\begin{figure*}
	\centering
	\includegraphics[width=\textwidth]{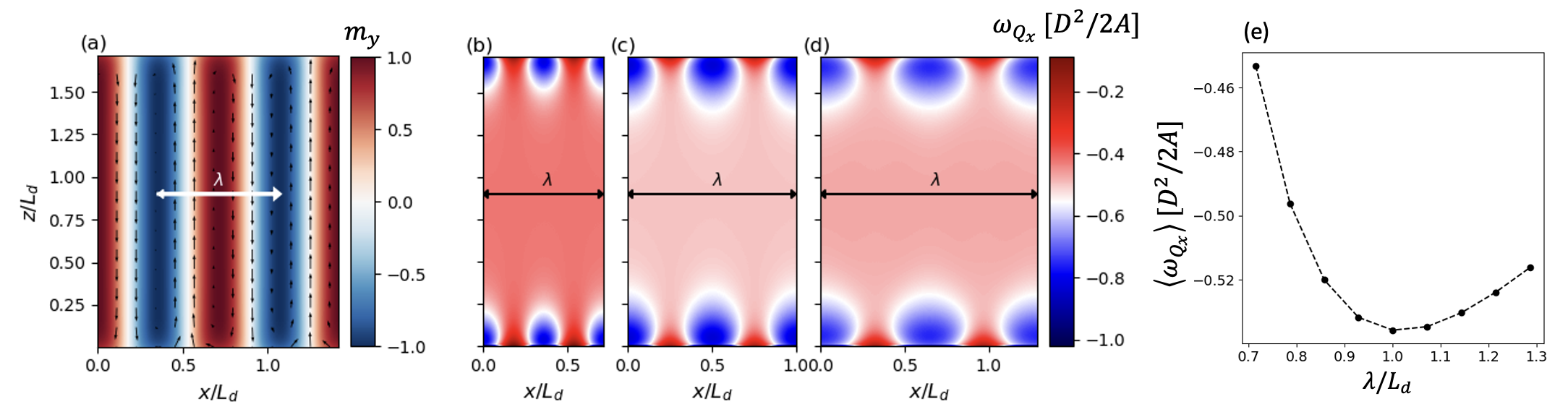}
	\caption{Effect of  wavelength on the energy of a zero-field $q_x$-helical state in an isotropic thin film of thickness $t = 1.714 L_d$. A representation of the helical texture is shown in (a), while subplots (b)-(d) show the micromagnetic energy density $\omega_{Q_x}$ for helical states with wavelengths of $\lambda = 0.714 L_d, 1.00 L_d$ and $ 1.29 L_d$ respectively. In (e), the average energy density of the helical state is shown as a function of helical wavelength, $\lambda$. }
	\label{wavelength_optimization}
\end{figure*}

\noindent
The equilibrium phase diagrams generated in this work are obtained via numerical minimization of Equation \ref{hamiltonian} using a finite-difference discretization of the Landau-Lifshitz-Gilbert equation (without the precession term), as implemented in the MuMax3 software package\cite{Vansteenkiste2014}. Periodic boundary conditions are applied along the $x$ and $y$ directions in order to capture the properties of a infinite thin-film. At each point in $(H, t)$ space, the optimized magnetic configurations are obtained by relaxing an approximation of the phase of interest (e.g. obtaining the cone+twist state by relaxing a regular conical state), as has been done previously\cite{Jin2017}. For phases that are periodic with some period $\lambda$, e.g. helical states with a certain wavelength or skyrmion lattice states with a certain lattice parameter, these minimizations are performed for states with a range of different $\lambda$ values in order to ensure that the true lowest energy state is found. An example of this optimization is shown in Fig. \ref{wavelength_optimization}, wherein a wavelength of approximately 1.0 $L_d$ is found to minimize the energy of the $q_x$-spiral state at zero field. \newline 

\noindent	
Cubic micromagnetic cells with lengths less than 0.01 $L_d$ are used when resolving the magnetic textures considered in this work. When generating the phase diagrams, the film thicknesses/applied field strengths are sampled in steps of approximately 0.07 $L_d$ and 0.012 $H_d$ respectively. After the minimal energies of each phase have been resolved onto this initial $(H, t)$ grid, the energies are then interpolated onto a finer grid using a 2D cubic spline interpolation. 

\section{Isotropic/Easy-Axis Films ($k_u \geq 0 $)}

\begin{figure*}
	\centering
	\includegraphics[width=\textwidth]{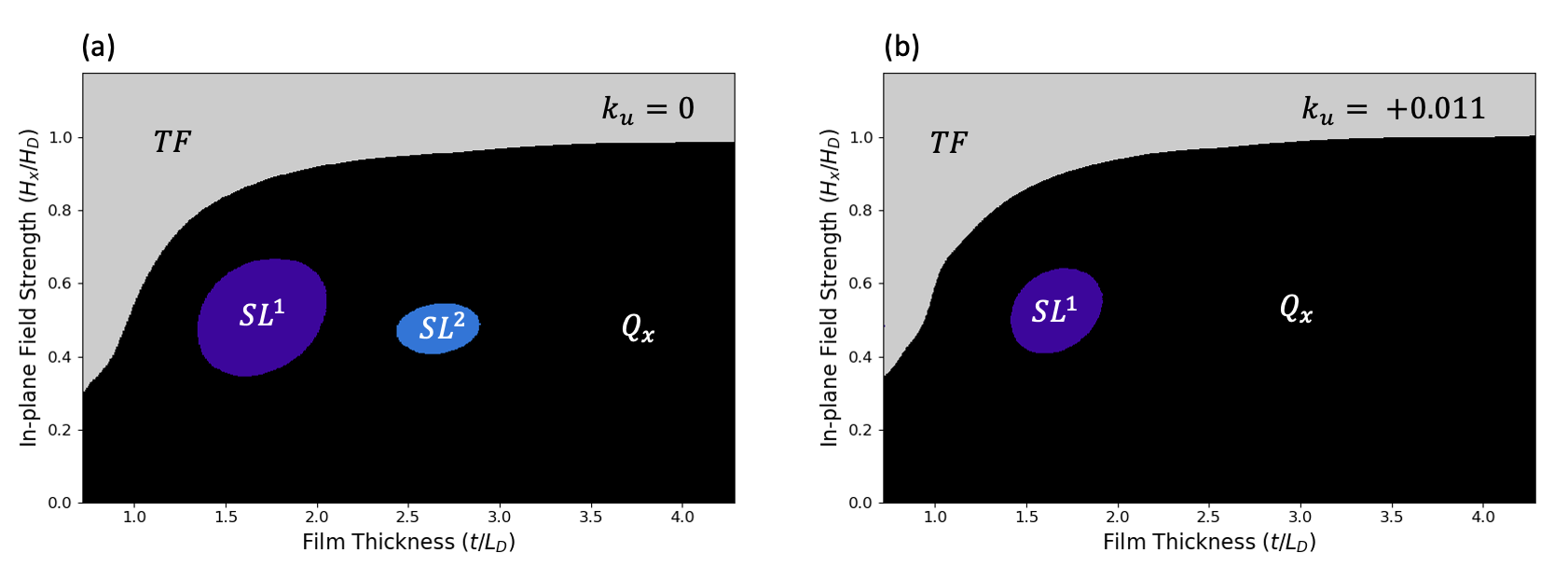}
	\caption{ Equilibrium magnetic phase diagrams for infinite thin-film chiral ferromagnets with unaxial anisotropy values of (a) $k_u=0$ and (b) $k_u = +0.011$. The labeled regions correspond to where the twisted ferromagnetic phase (TF), the $q_x$-conical phase ($Q_x$) and the one-layer and two-layer in-plane skyrmion lattice phases ($SL^1$ and $SL^2$, respectively) are the lowest energy phases.}
	\label{isotropic_PD}
\end{figure*}

\begin{figure*}
	\centering
	\includegraphics[width=\textwidth]{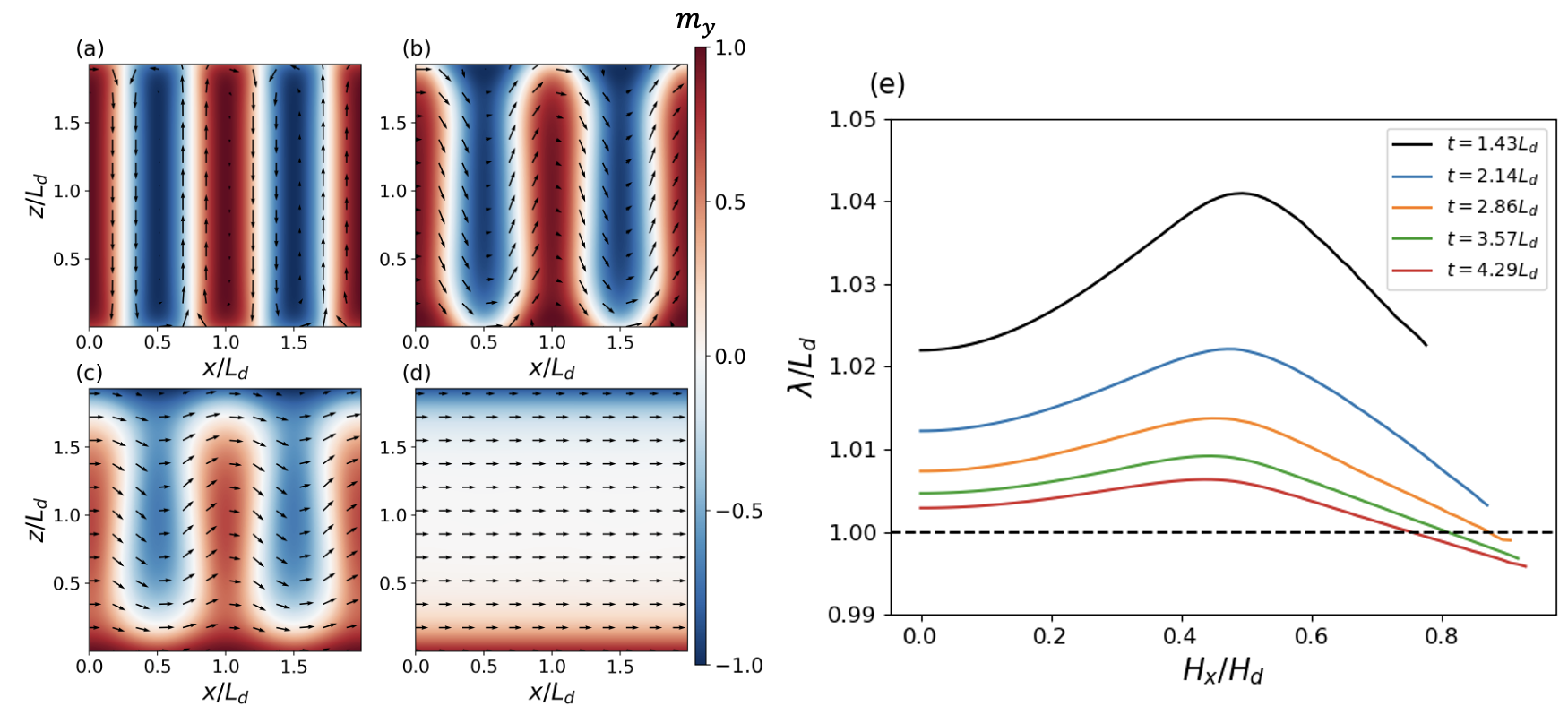}
	\caption{Evolution of the $q_x$-conical phase in increasing field for an isotropic film. Subplots (a)-(c) show the conical state in applied fields of $H = 0 H_d, 0.353 H_d$ and $0.706 H_d$, while (d) shows the twisted ferromagnetic phase in an applied field of $H=1.06 H_d$. In (e), the optimal conical wavelengths for different film thicknesses are plotted as a function of applied field, up to the field at which they transition into the twisted ferromagnetic state. }
	\label{qx_cone} 
\end{figure*}

\noindent
The equlibrium phase diagram for isotropic thin films is shown in Fig \ref{isotropic_PD}(a). For sufficiently large magnetic fields the equilibrium magnetic phase is a twisted ferromagnetic state where the system's magnetization is mostly oriented along the applied field direction, with the exception of a small amount of DMI-induced twisting near the film surfaces\cite{Wilson2013,Meynell2014b}. At low fields, the ground state is a helical spin texture with a helical vector oriented along the applied field direction (in this case $\hat{x}$). This phase, whose evolution in increasing field strength is shown in Figs. \ref{qx_cone}(a)-\ref{qx_cone}(d), is the thin film analog of the bulk conical state. It is sometimes refered to as the `cone + twist' state\cite{Jin2017}, in reference to the additional twists the conical texture acquires near the film surfaces. Throughout this work, however, we shall refer to it as the $q_x$-cone state, or simply the $q_x$ state or conical state. \newline 

\noindent
Although the surface twists do not have much of an effect on the optimal conical wavelength, as shown by the weak field and thickness dependence in Fig. \ref{qx_cone}(e), they do serve to lower the average energy density of the conical phase with respect to its bulk counterpart. The magnitude of this reduction in the average energy density goes approximately as the inverse of the film thickness $t$, such that the $q_x$-conical phase has a larger average energy density in thicker films. \newline 

\begin{figure}
	\centering
	\includegraphics[width=\columnwidth]{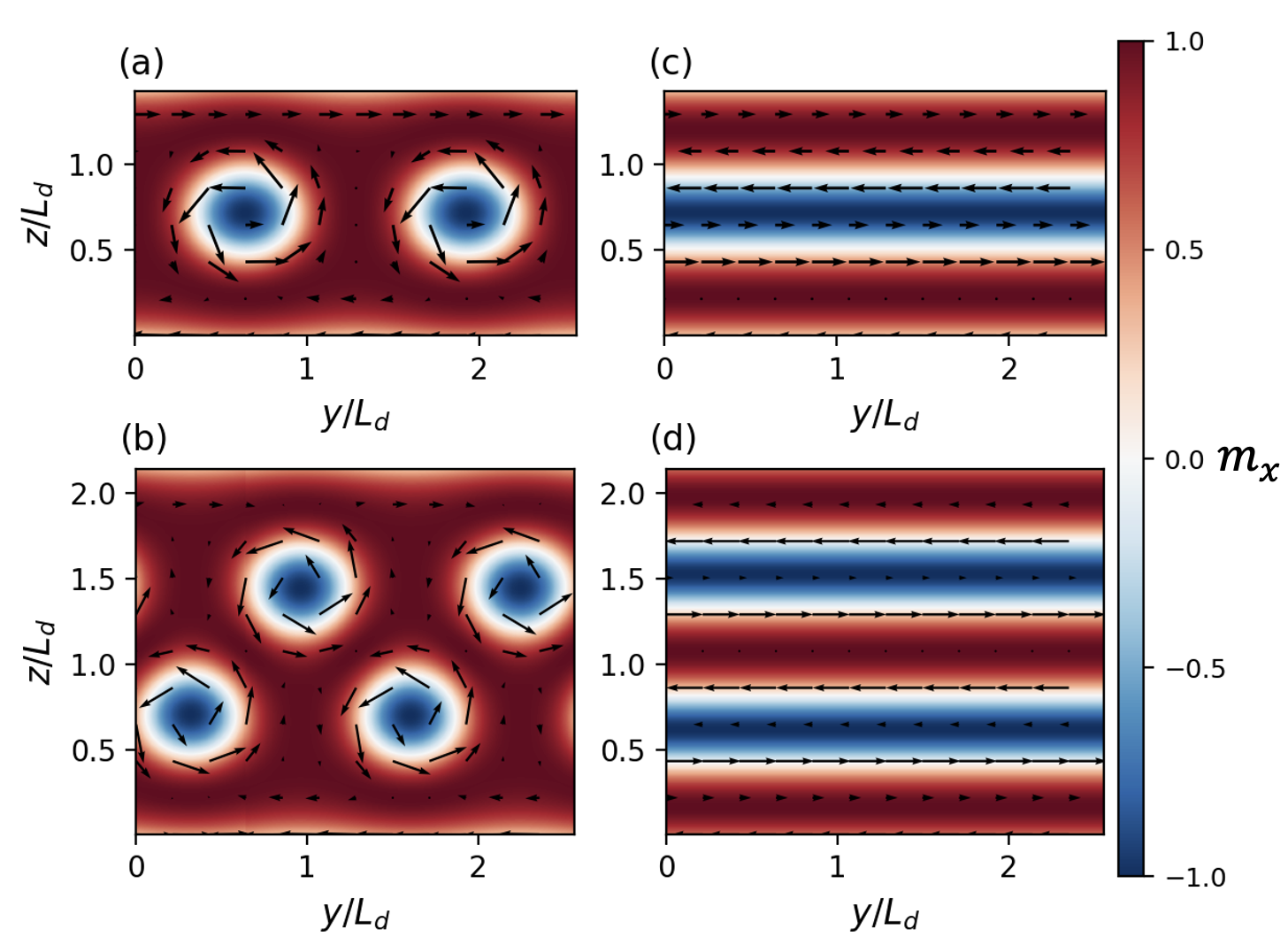}
	\caption{Representations of magnetic phases discretized by the geometric confinement along the $\hat{z}$ direction in an isotropic film. Subplots (a)-(b) show the one-layer and two-layer in-plane skyrmion lattice phases ($SL^1$, and $SL^2$, respectively) in an applied field of $H = 0.576 H_d$. Subplots (c)-(d) show $q_z$-helicoid states containing one and two helical windings ($Q_z^1$ and $Q_z^2$, respectively), in an applied field of $H = 0.235 H_d$. }
	\label{discrete_phases}
\end{figure}

\begin{figure*}
	\centering
	\includegraphics[width=\textwidth]{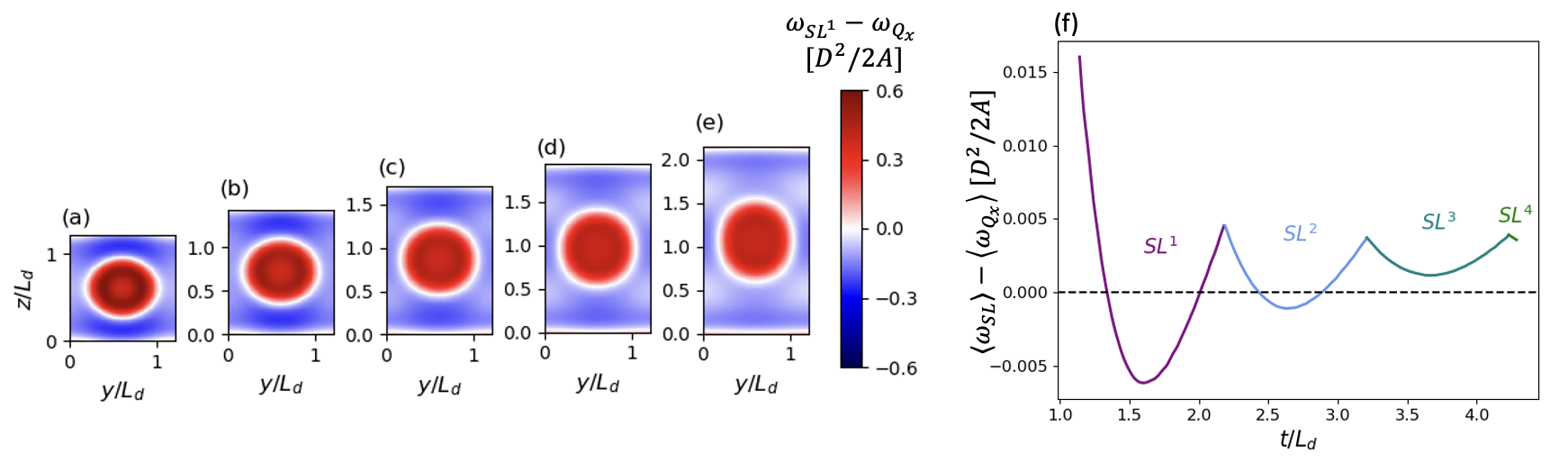}
	\caption{Relative energies of in-plane skyrmion lattices with respect to the average energy density of the $q_x$-conical phase in isotropic films. Subplots (a)-(e) show the relative energy density of the skyrmion lattice phase in films with thicknesses of $t = 1.21 L_d, 1.43 L_d, 1.71 L_d, 1.93 L_d$ and  $2.14 L_d$ respectively, in an applied field of $H=0.47 H_d$. In (f), the relative average energy density is given for one-layer, two-layer, three-layer and four-layer skyrmion lattices as a function of film thickness. }
	\label{1L_energies}
\end{figure*}

\noindent 
At intermediate field values, however, there exist discrete pockets in phase space where the equilibrium phase is either a single-layer or double-layer skyrmion lattice, depending on the thickness of the film. Representations of these magnetic phases are shown in Figs. \ref{discrete_phases}(a) and \ref{discrete_phases}(b). Figs. \ref{1L_energies}(a)-\ref{1L_energies}(e) show how the energy density of single-layer skyrmion lattice states compare to the average energy density of the $q_x$-cone state at the same applied field strength and film thickness. We see that while the skyrmion cores are themselves energetically unfavorable, their presence induces a large amount of chiral twisting in the surrounding regions, and especially between the cores and the film surfaces. Whether or not these surface twists are sufficient to stabilize the skyrmion lattice state depends on the thickness of the film. \newline 

\noindent 
In thinner films, the surface twisting takes place over a shorter (and more optimal) length scale, lowering the energy of the surface twists. However, at the same time, the energetically unfavorable core regions make up a larger fraction of the cross-section. The competition between these two trends causes the energy density of the single-layer skyrmion lattice state to be minimized for some critical thickness value, and it is only for a range of thicknesses in the neighborhood of this critical value that the single-layer state is the magnetic ground state, as can be seen in Fig. \ref{1L_energies}(f). This optimization problem is analogous to what occurs when varying the skyrmion-skyrmion separation, i.e., the skyrmion lattice parameter. \newline

\noindent
For sufficiently thick films, it becomes energetically favorable to have multiple layers of skyrmions instead of just one. For the two-layer skyrmion lattice state, an analogous minimum in the energy density occurs at a slightly larger film thickness. It is worth noting, however, that in the optimized magnetic textures, the regions between the skyrmions and the film surfaces are lower in energy than the regions between two skyrmions. As a consequence of this fact, the two-layer skyrmion lattice state has a larger average energy density than its single-layer counter part, since its surface twists make up a smaller fraction of the overall magnetic texture. This reduces the range of fields and thicknesses for which the two-layer skyrmion phase is the magnetic ground state, resulting in a smaller pocket in $(H, t)$ phase space. In the case of skyrmion lattices with three or more layers, for which the surface twists make up an even smaller fraction of the magnetic texture, the energy density is \textit{never} lower than that of the $q_x$-conical state (for isotropic films). This trend is consistent with the well-known fact that Equation \ref{hamiltonian} does not support a stable skyrmion phase in the bulk isotropic limit\cite{Muhlbauer2009}. In the absence of thermal fluctuations, surface effects are needed to stabilize the in-plane skyrmion lattice phases, and it is only for relatively thin films that these effects are strong enough to do so. \newline

\noindent
It is interesting to note that there is a range of film thicknesses between the $SL^1$ and $SL^2$ phase pockets for which skyrmion lattice phases are not stable for any value of the applied field $H$. This is a consequence of the system being infinite along the $\hat{x}$ direction. For chiral ferromagnetic systems that are finite along the applied field direction, the skyrmion tubes acquire surface twists where they meet the film edges, lowering their energy and enlarging their corresponding phase pockets\cite{Jin2017}. For sufficiently narrow systems, stable skyrmion phases can be obtained for all sufficiently thick films. Our results show that if the films grown are sufficiently wide (such that edge effects have only a minor effect on the overall energy density of the phases), then there are only two discontinuous ranges of film thicknesses for which isotropic films can support skyrmion phases. This fact may explain the absence of an in-plane skyrmion lattice phase in thick films of MnSi \cite{Wiedemann2017} (at low temperature), but does not explain its apparent absence in thin-films of FeGe\cite{Kanazawa2016}. \newline 

\noindent
When grown as epitaxial thin films, these materials can also acquire a uniaxial anisotropy along the direction of the film normal (i.e., $\hat{z}$) due to the strain induced by the mismatch between the chiral ferromagnet and the substrate on which it is grown\cite{Karhu2012}. In principle, the sign of this anisotropy depends on whether the substrate-induced strain is tensile or compressive. The effect of a weak easy-axis anisotropy on the magnetic phase space is shown in Fig. \ref{isotropic_PD}(b). The uniaxial anisotropy serves to lower the relative energy of the $q_x$-cone phase, which contains more spins with significant $m_z$ components than either the twisted ferromagnetic phase or the in-plane skyrmion lattices. This causes the single-layer skyrmion lattice pocket to shrink, and the double-layer pocket to be suppressed entirely. Stabilizing interesting in-plane phases, therefore, is more favorable with an easy-\textit{plane} anisotropy, the effects of which we study in the following section. 
\newline

\section{Easy-Plane Films ($k_u < 0$)}

\begin{figure*}
	\includegraphics[width=\textwidth]{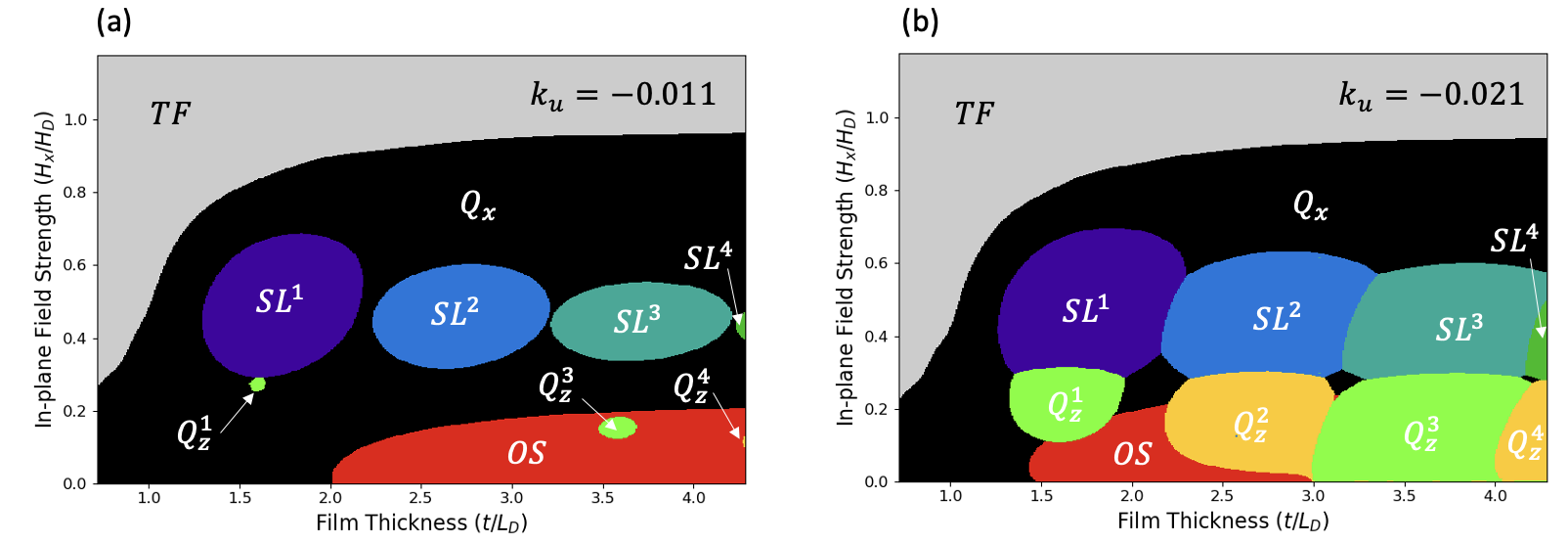}
	\caption{Equilibrium magnetic phase diagrams for infinite thin-film chiral ferromagnets with easy-plane uniaxial anisotropy values of (a) $k_u = -0.011$ and (b) $k_u = -0.021$. In addition to the phases introduced in Fig. \ref{isotropic_PD}, these diagrams also include the oblique spiral phase ($OS$), the three-layer and four-layer in-plane skyrmion lattices ($SL^3, SL^4$) and the discrete $q_z$-helicoid states ($Q_z^n$). }
	\label{aniso_PD}
\end{figure*}

\noindent
The effects of a small easy-plane anisotropy on the equilibrium phase diagram are shown in Fig. \ref{aniso_PD}(a). The main effect of this anisotropy is to increase the energy of the $q_x$-conical state, which contains a large number of spins with large $m_z$ components. This results in the increased stability of a number of different phases, which are relatively unaffected by the presence of an easy-plane anisotropy. Firstly, the one-layer and two-layer skyrmion phase pockets have enlarged, and there is now a range of thicknesses for which triple-layer ($SL^3$) and quadruple layer ($SL^4$) skyrmion lattice states become stable. \newline

\begin{figure*}
	\includegraphics[width=\textwidth]{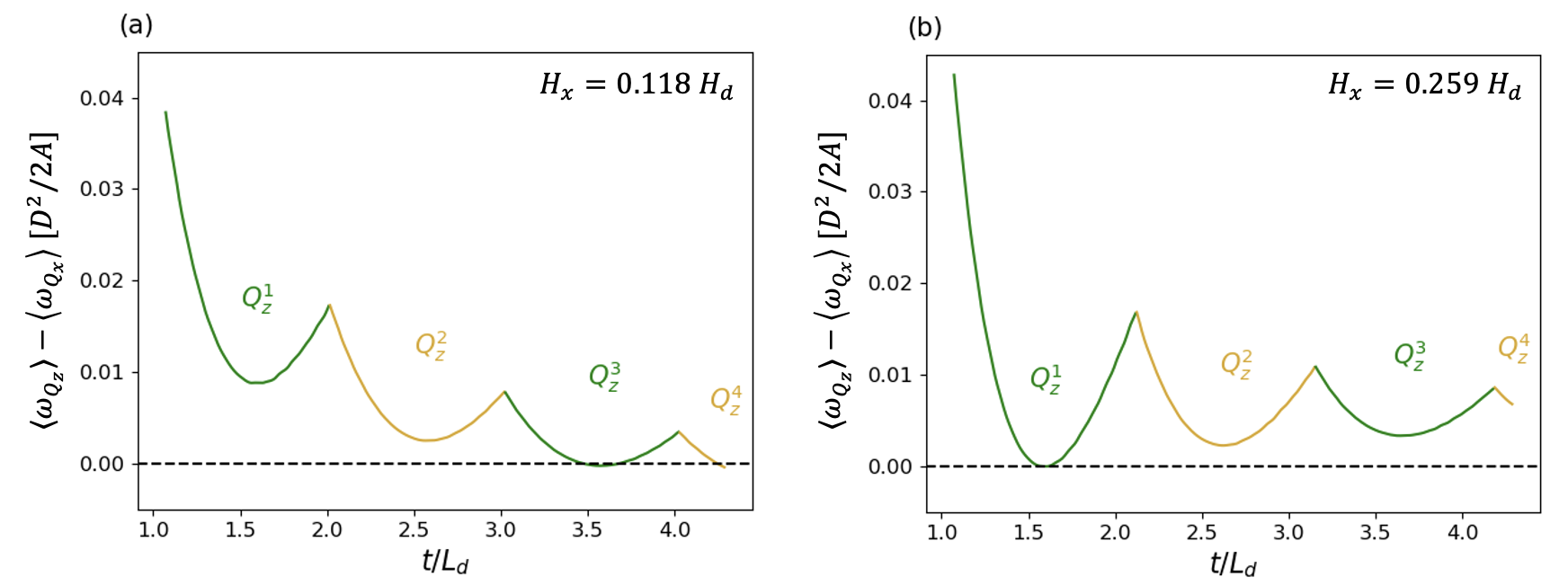}
	\caption{Relative energies of the discrete $Q_z^n$ helicoid states with respect to the $q_x$-conical phase as a function of thickness for applied fields of (a) $H = 0.118 H_d$ and (b) $H = 0.259 H_d$. A uniaxial anisotropy value of $k_u = -0.011$ is used for all thickness/fields. }
	\label{qz_energies}
\end{figure*}

\begin{figure*}
	\includegraphics[width=\textwidth]{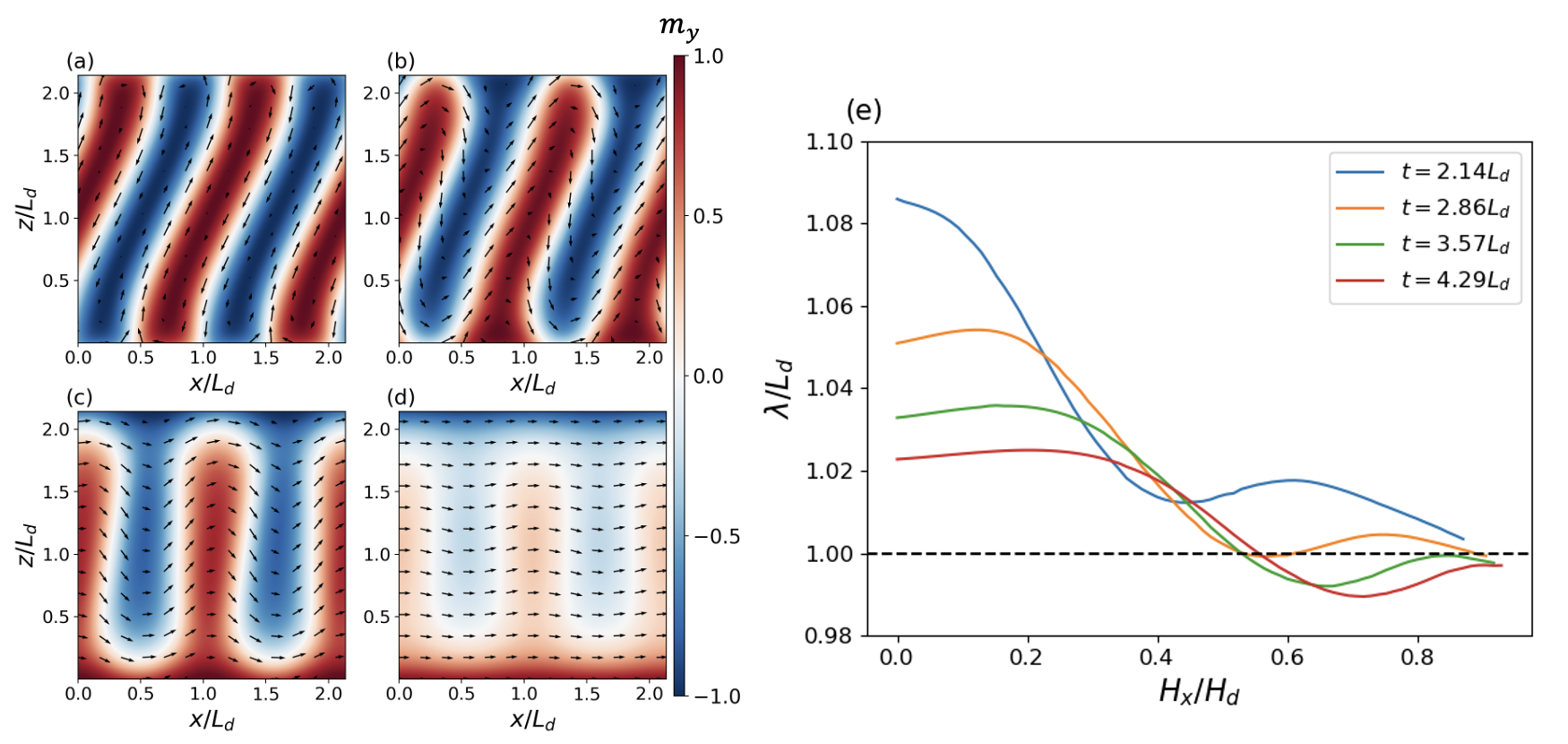}
	\caption{Evolution of the oblique spiral phase in increasing field. Subplots (a)-(c) show the oblique spiral phase in applied fields of $H = 0 H_d, 0.294 H_d$ and $0.588 H_d$, while (d) shows the $q_x$-conical phase at an applied field of $0.882 H_d$. In (e), the optimal wavelengths for different film thicknesses are plotted as a function of applied field. A uniaxial anisotropy value of $k_u = -0.011$ is used for all thickness/fields.}
	\label{oblique_spiral}
\end{figure*}

\noindent
Secondly, there are now small phase pockets corresponding to different $q_z$-helicoid states, i.e., discrete helicoidal states with propagation vectors along the out-of-plane direction.  At zero field, the wavelength of these states does not depend on the film thickness, and so the total amount of helicoidal twisting varies continuously as a function of film thickness. In the presence of a non-zero magnetic field, however, these states can minimize their Zeeman energy by adopting configurations containing a half-integer number of twists\cite{Wilson2013}, as shown in Figs. \ref{discrete_phases}(c)-\ref{discrete_phases}(d), hence maximizing the number of spins with positive $m_x$ components. This gives these states a quantized nature, with the amount of twisting changing discontinuously as a function of increasing film thickness. \newline

\noindent
Similar to the skyrmion lattice states, which possess a discrete number of skyrmion layers, the energy density of these discrete helicoid states also possesses a series of local minima as a function of film thickness $t$, as shown in Fig. \ref{qz_energies}(a). These minima lie near the points $t = (n + 1/2)L_d$, the thicknesses for which the resulting helicoidal wavelengths are near-optimal. Deviations from these optimal thickness values cause the helicoidal states to adopt non-optimal wavelengths, increasing their energies. Unlike the in-plane skyrmion lattice phases, however, for which the single-layer lattice can be stabilized (for some thickness) at all fields where the higher-order skyrmion lattices are stable, the field-dependence of the $q_z$-helicoid states is more involved. At weaker fields, it is only the states with a large number of helicoidal windings that are lower in energy than the $q_x$-conical phase, as shown in Fig. \ref{qz_energies}(a),  while Fig. \ref{qz_energies}(b) shows that at intermediate fields this trend is reversed. This trend reversal results in the interesting topology shown in Fig. \ref{aniso_PD}(a), where there are stable phase pockets corresponding to the $Q_z^1, Q_z^3$ and $Q_z^4$ states, but \text{not} to the $Q_z^2$ state. \newline

\noindent
This behaviour can be understood in terms of the Zeeman interaction. At zero field, the $q_z$-helicoid state's energy density has no thickness dependence. The average energy density of the $q_x$-conical state, in contrast, increases as a function of sample thickness, due to the decreasing importance of the low-energy surface twists shown in Fig.  \ref{wavelength_optimization}. For sufficiently weak fields, therefore, the relative energy of the $Q_z^n$ helicoid states is lowest in thicker films. At intermediate fields, however, the Zeeman interaction becomes much more important. In the $Q_z^1$ helicoid state shown in Fig. \ref{discrete_phases}(c), approximately two-thirds of the texture's spins have positive $m_x$ components, resulting in a relatively large reduction in the system's energy as the applied field is increased. In thicker films, however, the average value of the $m_x$ components is much closer to 0, and the corresponding reduction in the system's energy at higher fields is greatly reduced.  \newline

\noindent
Lastly, for sufficiently thick films, the ground state at low field is now the so-called oblique spiral state\cite{Leonov2020b}, which can be thought of as a $q_x$-conical state that has canted itself slightly in order to minimize the average magnitude of its $m_z$ components. The evolution of an oblique spiral state in increasing field is shown in Figs. \ref{oblique_spiral}(a)-\ref{oblique_spiral}(d).  For the easy-plane anisotropy values considered in this work, the oblique spiral can persist as a metastable state outside of its equilibrium phase pocket, eventually transitioning into a regular $q_x$-cone phase at some thickness-dependent critical field. This continuous transition occurs at larger fields in thicker films. While in a sufficiently thick film, this canting of the $q_x$-conical phase lowers the micromagnetic energy everywhere within the texture, in thinner films the distortions induced by this canting become energetically unfavorable. There is therefore a critical thickness below which the oblique spiral phase is unstable. \newline 

\noindent
Figure \ref{aniso_PD}(b) shows the effect of a slightly stronger easy-plane anisotropy on the equilibrium phase diagram. The $q_x$-conical state has been even further suppressed, and as a result the skyrmion phase pockets have expanded to the point that all sufficiently thick films now support a stable skyrmion phase. The $Q_z^n$ phase pockets have also expanded considerably, including that of the previously absent $Q_z^2$ phase. Interestingly, there remains a range of thicknesses between the $Q_z^1$ and $Q_z^2$ pockets for which the discrete in-plane helicoid is never the equilibrium ground state. This feature can be understood in terms of the geometric confinement of the helicoidal states. In sufficiently thin films, the $q_z$-helicoid state contains only a single helicoidal twist. Any change in the thickness of the film, therefore, will produce an similarly large change in the wavelength of this single twist, and hence in the phase's energy density. In thick films, however, the effect of a change in the film thickness will be spread out over a large number of helicoidal twists, and the corresponding change in wavelength of each individual twist will be significantly reduced. We therefore find that in thick films, the energy density of the $q_z$-helicoid state is relatively insensitive to the film thickness. In thin films, however, deviations from the optimal thickness for the $Q_z^1$ phase produce large increases in the average energy. This effect is clearly shown in Fig. \ref{qz_energies} (b).  \newline

\noindent
It is worth noting that while there is a large and continuous range of thicknesses for which the ground state at low field is one of the $q_z$-helicoid states, the range of thicknesses for which one can expect to observe a transition between helicoids with different winding numbers (e.g. $Q_x^3 \to Q_x^2$) upon varying the field strength remains small and discontinuous. Nevertheless, further experimental exploration of these transitions would be worthwhile, as the results obtained may offer some insight into the origin of the qualitative difference in behavior observed in MnSi\cite{Wilson2013}, where the transition is abrupt, and in FeGe\cite{Kanazawa2016}, where the transition occurs via an intermediate incommensurate phase. \newline

\noindent
Lastly, while at larger film thicknesses the oblique spiral phase has been suppressed in favor of the discrete $Q_z^n$ helicoids (for which $m_z$=0 everywhere), it remains stable at lower thicknesses, and the critical thickness below which it becomes unstable has been reduced. Nevertheless, we see that the range of thicknesses over which the oblique spiral state can be stabilized has decreased in magnitude. As a consequence, there is only a relatively narrow range of anisotropy values $k_u$ that are ideal for experimental investigations of the oblique spiral phase in these thin-film systems. \newline

\section{Conclusions}

\noindent
In summary, we have investigated the equilibrium phase space of magnetic textures in infinite thin-films of chiral ferromagnets, focusing on the effect of in-plane magnetic field strength $H$, film thickness $t$, and easy-plane uniaxial magnetocrystaline anisotropy $K_u$. We find that the absence of edge effects (not to be confused with surface effects) results in a phase space with a significantly different topology than has been previously calculated for nanostripe geometries. Notably, the range of film thicknesses which support stable in-plane skyrmion lattices and/or stable in-plane helicoid phases are found to be disconnected in the $(H,t)$ parameter space for small values of the uniaxial anisotropy, suggesting an interesting direction for future experiments. In particular, we find that systems with certain easy-plane anisotropy values can support $q_z$-helicoid states with one or three twists, but not two. Furthermore, we find that in weak fields only sufficiently thick films can support these helicoidal states, while at intermediate fields the films must be sufficiently \textit{thin}.   We explain these features in terms of the interplay between the applied field and the geometric confinement provided by the film surfaces. \newline

\noindent More generally, our results can be used as a road map for future investigations of thin-film chiral ferromagnetic systems, guiding researchers towards the most promising regions of $(H, t, k_u)$ phase space for the observation of certain phases (e.g. the oblique spiral phase) or certain field-induced phase transitions (e.g. the unwinding of $Q_z^n$-helicoid states).

\begin{acknowledgments}
The work of C. R., M. L. P. and T. L. M. was supported by the Natural Sciences and Engineering Research Council of Canada (NSERC), and by resources provided by ACENET (ace-net.ca) and the Digital Research Alliance of Canada (alliancecan.ca). C. R. acknowledges support from NSERC PGS-D and Nova Scotia Graduate scholarships. 
\end{acknowledgments}
		
\bibliography{main}
		
\end{document}